# EDEFuzz: A Web API Fuzzer for Excessive Data Exposures

Lianglu Pan, Shaanan Cohney, Toby Murray, Van-Thuan Pham
lianglup@student.unimelb.edu.au,shaanan@cohney.info,{toby.murray,thuan.pham}@unimelb.edu.au
The University of Melbourne, Melbourne, Australia

## ABSTRACT

APIs often transmit far more data to client applications than they need, and in the context of web applications, often do so over public channels. This issue, termed *Excessive Data Exposure* (EDE), was OWASP's third most significant API vulnerability of 2019. However, there are few automated tools—either in research or industry—to effectively find and remediate such issues. This is unsurprising as the problem lacks an explicit test oracle: the vulnerability does not manifest through explicit abnormal behaviours (e.g., program crashes or memory access violations).

In this work, we develop a metamorphic relation to tackle that challenge and build the first fuzzing tool—that we call EDEFuzz—to systematically detect EDEs. EDEFuzz can significantly reduce false negatives that occur during manual inspection and ad-hoc text-matching techniques, the current most-used approaches.

We tested EDEFuzz against the sixty-nine applicable targets from the Alexa Top-200 and found 33,365 potential leaks—illustrating our tool's broad applicability and scalability. In a more-tightly controlled experiment of eight popular websites in Australia, EDEFuzz achieved a high true positive rate of 98.65% with minimal configuration, illustrating our tool's accuracy and efficiency.

**ACM Reference Format:**
Lianglu Pan, Shaanan Cohney, Toby Murray, Van-Thuan Pham. 2024. EDEFuzz: A Web API Fuzzer for Excessive Data Exposures. In *2024 IEEE/ACM 46th International Conference on Software Engineering (ICSE 2024), April 14–20, 2024, Lisbon, Portugal*. ACM, New York, NY, USA, 12 pages. https://doi.org/10.1145/3597503.3608133

## 1 INTRODUCTION

Every week, another leak! Server-side APIs of web applications frequently transmit more data than is needed for their corresponding clients. This may not have been an issue, were it not for the fact that these APIs are often publicly accessible. API vulnerabilities of this type are known as *Excessive Data Exposures* (EDEs). Despite ranking as OWASP's #3 most significant API vulnerability for 2019 [1], technology to detect these vulnerabilities remains underdeveloped.

We thus develop the first automated and systematic fuzzing tool, EDEFuzz, to detect EDEs. As the "gold standard for finding and removing costly, exploitable security flaws", fuzzing is a key tool for cost-effectively detecting and remediating such issues [2].

We posit that the lack of automated tools to detect EDEs is due to their *semantic* nature. Specifically, EDEs do not manifest through explicit, abnormal behaviours (e.g., program crashes). Detecting them thus requires a model of what constitutes an EDE.

> "Automatic tools usually can't detect this type of vulnerability because *it's hard to differentiate between legitimate data returned from the API, and sensitive data* that should not be returned without a deep understanding of the application."
> — The Open Web Application Security Project (OWASP)

> "This vulnerability is so prevalent (place 3 in the top 10) because it's easy to miss. *Automation is near useless here because robots can not tell what data should not be served to the user without telling them exactly how the application should work.* This is bad because API's are often implemented in a generic way, returning all data and expecting the front-end to filter it out."
> — Wallarm End-to-End API Security Solution

**Figure 1: Industry views on the EDEs. These indicate the prevalence of EDEs and limitations of existing detection tools**

We start with a definition: an API is vulnerable to EDE if it exposes meaningfully more data than what the client legitimately needs [1].

Consider a simple example of an online storefront. When a user views the page for a specific product, an API call may be made to fetch stock levels, informing the user whether the item is in stock. However, the API may also return extraneous data (such as the profit margin on the item) that is not displayed to the user but is nonetheless transmitted. The transmission of the extra data constitutes an "excessive data exposure". This leads to our motivating question:

*How to detect if a web API exposes unnecessary data?*

The question is related to the famous test oracle problem. How can a tester or an automated testing technique distinguish desired, correct behaviour from undesired or incorrect behaviour [3]. The common wisdom in industry (see **Figure 1**) is that the test oracle problem renders EDE detection beyond current testing approaches.

We address this challenge with the following key insight:

*Data returned from an API endpoint is more likely excessive if it has no impact on the content displayed to a user.*

Specifically, we develop the following novel metamorphic relation[1] to side-step this problem. Through the relation, automated testing approaches can check if a data field in an API response is excessive by checking for difference between what a client displays when the field is present in an API response, versus when the field is deleted.

Formally, assume we have an API response under analysis $\mathcal{R}_{\text{origin}}$ comprising a set of data fields. A web client (e.g., a web browser) uses $\mathcal{R}_{\text{origin}}$ to render a page that can be represented by a Document Object Model (DOM) tree $\mathcal{D}_{\text{origin}}$. A data field $d \in \mathcal{R}_{\text{origin}}$ is considered non-excessive if the following inequality holds:

$$\text{diff}_{\text{DOM}}(\mathcal{D}_{\text{origin}}, \mathcal{D}_{\text{mutated}}) \neq 0, \tag{1}$$

---
[1]A metamorphic relation is one that holds between two different program inputs and their corresponding outputs [4]

where diff_DOM calculates the difference between two DOM trees $\mathcal{D}_{origin}$ and $\mathcal{D}_{mutated}$. $\mathcal{D}_{mutated}$ is constructed from $\mathcal{R}_{mutated}$ which we obtain by removing the in-question data field $d$ from $\mathcal{R}_{origin}$. If a data field violates **Equation (1)**, it is deemed excessive.

From this relation we build a system that significantly reduces the potential for false negatives that hinder competing approaches—manual inspection and keyword-matching. Notably, keyword matching techniques often use a list of terms ("key", "token", "password" etc) in order to flag exposures [5] and therefore any excessive data field that does not match any known keywords is erroneously ignored.

In contrast to these approaches, our tool EDEFuzz leverages the metamorphic relation to detect EDEs. It mutates and replays API responses into the client side of a web application and compares the generated DOM tree with the original tree in each fuzzing iteration.

Building the tool required surmounting two major challenges.

First, we needed to build an API fuzzer with *response determinism*. Existing mutation algorithms used in Web API testing/fuzzing [6] focus on mutating API *requests* which introduces random and untargeted changes in server *responses*. However, our metamorphic relation requires that the responses differ only in a single field.

Second, like other fuzzing tools, the usefulness of our tool depends on its ability to achieve reasonable throughput (represented in tests per second). This challenge is particularly acute in the context of web fuzzing as tools are rate limited by both bandwidth, and server load. For public sites, the challenge is further compounded by both server-side rate-limiting and the need to minimize disruption. These hinder the timely progress of a fuzzing tool.

To address these two challenges, we adopt a "record-replay" model [7]. We combine a web proxy and a custom-built simulated server to minimize interactions with sites under-test. Prior to beginning the fuzzing process, our tool initiates a "record" phase: a web proxy captures all client requests and server responses, including the request sent to the targeted API and the corresponding response. Note that in each fuzzing campaign EDEFuzz targets only one API. Following the record phase, fuzzing begins (i.e., the "replay" phase).

In the replay phase, *no communication with the actual remote server is necessary*. Our lightweight simulated server handles all requests. If a request is sent to the targeted API, the simulated server transmits a mutated version of the original server response. Otherwise, the simulated server merely replays the recorded transmissions.

This architecture yields several benefits. First, test executions (i.e., sending requests and getting responses) are performed locally—leading to much lower latency. Second, changes to the remote server do not impact test results, making them highly deterministic. Maintaining deterministic results is a critical requirement for fuzzing in general because it helps reduce false positives. However, when detecting EDEs, this also helps reduce false negatives. Absent this determinism, an application change that yields a different web page may cause EDEFuzz to incorrectly flag a field as non-excessive–believe the DOM change to be caused by changes in the server-response and not in the application itself. Third, the architecture permits running tests in parallel, which minimizes the burden of scaling the tool.

We evaluate the tool in two different settings. First motivated by a recent massive Web API leak in Australia [8], we applied EDEFuzz to several comparable Australian web properties. We perform a detailed comparison of the tool's results against a corresponding manual effort to assess the severity and accuracy of the findings. Second, we run our tool against a broader set of sixty-nine web applications—the complete set of applicable targets from the Alexa Top-200. We use this evaluation to assess the scalability of our tool as well as its applicability to a representative set of global web applications.

Our overall contributions are as follows:
- We identify a novel metamorphic relation to address the test oracle problem in the context of detecting excessive data exposure.
- We develop the first systematic and automated fuzzing tool for detecting excessive data exposure vulnerabilities, EDEFuzz.
- We empirically evaluate the accuracy of our approach, its applicability to popular websites, and its efficiency (both in terms of computational time and human effort). Our results demonstrate EDEFuzz's effectiveness for discovering unknown sensitive data leakage via EDE also, whose prevalence we also investigate. We found that our approach is
  – highly accurate: 98.65% of the fields flagged by the tool in a controlled study were true excessive data exposures.
  – widely applicable to popular websites, requiring modest computational costs and human effort to employ.
  – able to discover zero-day EDE vulnerabilities. Specifically, it found five zero-day EDE vulnerabilities serious enough to merit immediate disclosure.

To support future research in this interesting topic, we will make EDEFuzz open sourced. The source code, along with some tutorials will be available at https://github.com/Broken-Assumptions/EDEFuzz.

In **Section 2**, we provide the necessary preliminaries on Web APIs, Excessive Data Exposures and Metamorphic Fuzzing. In **Section 3**, we motivate our work with several real-world vulnerabilities *discovered by our tool*. In **Section 4**, we present our automated approach to detect EDEs and our implementation. In **Section 5**, we report our experimental results. We consider related work in **Section 6** followed by a brief discussion in **Section 7**.

## 1.1 Research Ethics

We considered both the propriety of our scanning and fuzzing techniques and engaged in vulnerability disclosure.

We discussed our research in a series of conversations with our research ethics office who ultimately deemed it exempt from a full review process. Our research involves scraping and scanning commensurate with ordinary activity by both search engines and the research community. Our methodology minimizes interaction with remote servers by performing all fuzzing offline on a simulated replica of the target server. Given the low impact of capturing the outcome of a limited number of HTTP requests and the potential benefits of our research it was determined that our work adheres to the principle of beneficence that is the hallmark of research ethics.

As our work notes, EDEFuzz flags fields for further human analysis (rather than indicating vulnerabilities with certainty). As a result, our assessment of whether a flagged field rises to a reportable level requires human judgement about whether an EDE leaks sensitive information and the potential harms from that leak. In the five instances where we discovered sensitive data leakage we contacted the affected entities. By time of submission, all five entities had acknowledged our disclosures; two had remediated the issues.

## 2 BACKGROUND
### 2.1 Web APIs & Excessive Data Exposure

Web applications often expose API endpoints to the public internet. Exposing the endpoint allows the application to separate front- and back-end logic. While the front-end components focus on rendering visual elements and their associated interactive components, back-end logic is more closely tied to long-term data storage. The API allows the font-end to query the back-end and in many cases serves a response in either JSON or XML. While an API ought to narrowly tailor the data served in a response to the request, this practice is often ignored. OWASP terms this excessive data exposure (EDE) [9].

One cause for EDEs is that API developers over-rely on API clients to perform data filtering. This eases the cognitive burden on back-end developers to determine the specific needs of the client *a priori*.

When present, EDEs are often trivial to exploit. To obtain the excess (or even sensitive) data, it is often sufficient to simply examine response traffic from the target API.

Since technology to scan for and detect EDEs remains underdeveloped, OWASP only provides general advice [9] on how to prevent them, such as "Never rely on the client to filter data!", "Review all API responses and adapt them to match what the API consumers really need", and "Enforce response checks to prevent accidental leaks of data or exceptions". However, the prevalence of EDEs shows that advice alone is insufficient. We need effective automation!

### 2.2 Fuzzing

Fuzzing is a process of repeatedly generating (random) inputs and feeding them to a system under test (SUT) discover bugs. In its traditional use, a fuzzer detects issues through aberrant program behaviour, such as program crashes. This indicates a potential security bug in the SUT. In response, the fuzzer will preserve the bug-triggering input for further analyses (e.g., manual debugging).

While we preserve the input generation phase above, our work notably deviates in that we detect potential errors through *the lack of change* in program output, rather than a spec violation or crash.

However, our work is not the first to address API testing. Web API fuzzing recently garnered increased interest from both industry and academia [6], [10], [11]. RESTler [6] is the current state-of-the-art approach. RESTler is a stateful black-box REpresentational State Transfer (REST) API fuzzing tool. For a target with an OpenAPI/Swagger specification [12], RESTler analyzes its entire specification, and then generates and executes tests through its REST APIs.

Researchers typically classify fuzzers based on the level of integration between the SUT and the fuzzer. The most common classification is based on the fuzzer's awareness of the internal structure of the SUT [13]. Specifically, a black-box fuzzer knows nothing about the internal structure of the SUT. In contrast, a white-box fuzzer would know everything about the SUT such as its source code, its control flows and data flows. Grey-box fuzzing is in between; it leverages partial knowledge of the SUT, typically via lightweight program instrumentation. Our tool EDEFuzz is a black-box fuzzer.

### 2.3 Metamorphic Testing/Fuzzing

We adopt fuzzing as our approach to detecting EDEs because of its demonstrated success in discovering security flaws.

Highlighting the effectiveness of fuzzing, as of May 2022, Google's fuzzing infrastructure had detected over 25,000 bugs [14]. However, these bugs were detected using explicit test oracles. Bugs with a test oracle either lead to program crashes (e.g., segmentation faults) or are caught by instrumentation-based checkers (e.g., Address Sanitizer, etc). In contrast, semantic bugs like EDEs do not manifest through explicit abnormal behaviours, and cannot be reliably detected by observing *single* program executions.

How do we build tools to detect semantic bugs? Metamorphic testing and fuzzing, which leverage metamorphic relations, are a promising approach. At its core, metamorphic fuzzing involves *comparing* multiple executions of the SUT under different inputs and observing whether some relation (called the *metamorphic relation*) holds between their corresponding outputs.

Consider the following toy example of metamorphic bug finding: we can test a function that reverses a list by testing, for an arbitrary input list $x$, whether reversing the reverse of $x$ yields $x$ itself; or for a function that calculates distance between a pair of points, whether the distance from point $a$ to $c$ is always smaller or equal to the sum of the distances from $a$ to $b$ and from $b$ to $c$, etc.

Metamorphic relations are properties that must necessarily hold with respect to the correct functioning of the SUT. In metamorphic testing, the violation of a metamorphic relation indicates a potential bug [4]. Past work has successfully identified and used metamorphic relations to find bugs in a variety of systems, including Machine Translation software [15] and popular DBMSs [16].

## 3 MOTIVATING EXAMPLES

In this section, we present a selection of real-world EDEs detected by our tool to demonstrate their prevalence and implications, the challenges in detecting them using prior approaches, and the motivation for our tool (which we comprehensively evaluate in **Section 5**).

**Vulnerability 1 - Locations and Contact Details** EDEFuzz discovered a vulnerability while testing the live delivery tracking service offered by Company-I, an Australian last-mile delivery service. The vulnerability has been reported and fixed. As shown in **Figure 2**, a customer receives a unique link on the day that an item is on board for delivery. The link opens a web page which displays the name and a photo of the delivery driver, an Estimated Time of Arrival of the delivery, and the position of the item in the driver's queue. The page sends an API request to the server regularly, and updates the contents on the page based on the API response. Part of this response is shown in **Listing 1**, with sensitive information removed.

The client-side logic allows the webapp to display the accurate geographic location of the delivery driver *only* when the item to be delivered is at the front of the queue. It suggests that while the driver is delivering an item to a customer, another customer should not be able to ascertain the location of the driver—which would leak the location of other deliveries. However, EDEFuzz detected that the API response always contains rich information about the delivery driver, including accurate latitude and longitude (the `location` field), direction of facing (the `bearing` field), and speed of travelling (the `speed` field). EDEFuzz also identified the driver's manager's information in the API response.

Knowing the timestamped location of the delivery driver, a customer may recover the route that the delivery driver is travelling, or even be able to identify the address of other customers who receive parcel from the same delivery driver. Such information is ideal for enabling further attacks, like social engineering [17].

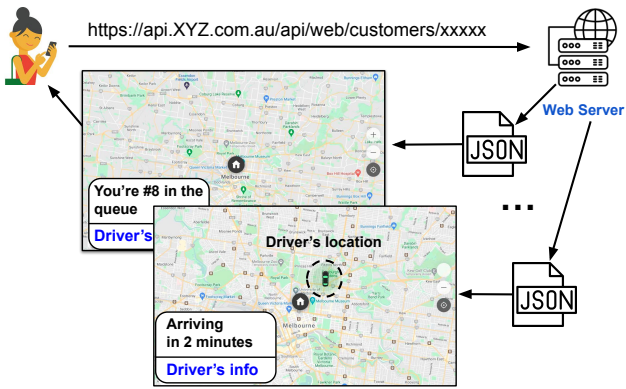

**Figure 2: API flow for a package delivery service. A web application requests tracking information, returned in a JSON object.**

**Listing 1: An API response to a query for delivery status. The authors have redacted or adjusted sensitive information.**

```
1  {
2    "driver":{
3      "id":353,
4      "url":"/api/web/drivers/353",
5      "full_name":"[NAME]",
6      "car":{
7        "car_type":1,
8        "car_type_name":"Car",
9        "capacity":33
10     },
11     "member_id":37270063,
12     "avatar":"[URL_TO_AVATAR]",
13     "is_online":true,
14     "work_status":"working",
15     "phone":"[PHONE_NUMBER]",
16     "status":"in_progress",
17     "location":{
18       "id":1081725,
19       "timestamp":1632107809.0,
20       "speed":7.60149761928202,
21       "accuracy":4.88680554160758,
22       "location":"-
          ↪  37.79998905537697,144.9940922248341",
23       "created_at":"2021-09-20T13:16:49.365492+10:00",
24       "bearing":277.508087158203,
25       ...
26     },
27     ...
28   },
29   "manager":{
30     "id":305,
31     "full_name":"[NAME]",
32     "email":"[EMAIL]",
33     "avatar":null,
34     "is_online":false,
35     "work_status":"not_working",
36     "phone":"[PHONE_NUMBER]",
37     "can_make_payment":false,
38     "merchant_position":"Manager",
39     "role":"Manager",
40     ...
41   },
42   ...
43 }
```

**Vulnerability 2 - Stock Levels** This second vulnerability exposed detailed stock availability for a retailer, which might be considered sensitive. Many retailers allow potential customers to check their stock availability online before visiting their shops. Some retailers reveal precise stock values on their websites, while other retailers decided to only display categorised values such as "in stock", "low stock" and "out of stock". Interestingly, we noticed a few such online services in which the server transmits the precise stock values but the web application only displays categorised values based on thresholds.

We observed this design in retailer companies such as Company-C, Company-D and Company-E. Company-D claimed that their accurate stock level is non-sensitive, whereas Company-E removed their stock level from their API response before we tried to contact them.

Regarding attack scenarios, if a retailer knows about the stock details of their competitors in different locations, they could adjust their logistic plan accordingly to increase their sales and gain more profit. Individual suppliers might also take advantage of this vulnerability to increase their prices for specific retailers when they know that they are low on stock. Having access to that kind of information, a 3rd-party company (e.g., a shopping suggestion service) could also develop an app and give more precise suggestions to their customer, leading to some monetary benefit.

**Vulnerability 3 - Network bandwidth** Transmitting excessive data fields reduces performance and thus the web application's user experience. Further, it imposes increased bandwidth requirements creating accessibility issues. EDEs imply a lack of consensus between an application's back- and front-ends. Like other "code smells" EDEs are indicative of poor development practices that may cause other vulnerabilities. We found a case in Company-C's API where the front-end consumed only 4.4% of the response data fields.

**The Bugs Escaped Detection!** There are a variety of ways in which these flaws may have escaped notice. Developers, while aware of security flaws, are heavily reliant on tooling to detect bugs. As discussed in Section 2 keyword-matching tools [5] to discover EDEs are of low effectiveness and prone to false negatives. To confirm this hypothesis we applied the popular vulnerability scanner Burp Suite [5] to attempt to identify EDE and sensitive data leakage to sites from our test set. As expected it failed to identity Vulnerability-2 in Company-D (a false negative); it also failed to identify almost all of the EDE and sensitive data leakage in Company-I (identifying only the disclosed email addresses in Listing 1). It also reported false positives (mistaking public postcode data for credit card numbers).

A tool-free approach requires back-end developers to carefully ascertain if a given field is needed, which is not always feasible.

As discussed in the case of Vulnerability-2, while Company-D claimed that their accurate stock level is non-sensitive, Company-E decided to remove that information from their API responses. This further indicates the need for a principles based approach that reduces the burden on developers to determine the balance between business need and sensitivity.

## 4 OUR APPROACH

We now provide an overview of EDEFuzz and detail the design of its components. We depict the main workflow of EDEFuzz when testing a website (given its URL and a targeted API endpoint) in Figure 3.

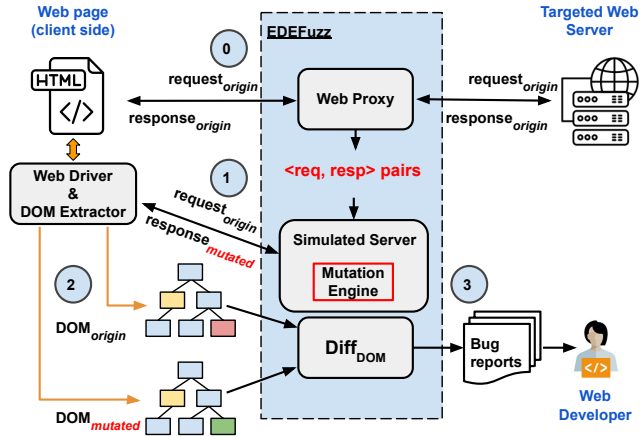

**Figure 3: The workflow of our approach to identify excessive data exposure vulnerabilities in web APIs.**

## 4.1 Overview

**Identifying API Endpoints.** Identifying target API endpoints is an orthogonal problem to testing them for the presence of EDEs, and is thus out of scope for this paper. One may leverage tools like crawlers to automatically explore the website and detect APIs for targeting. However, this automated approach could be destructive and potentially illegal without permission from the website owners. In our experiments (as detailed in **Section 5**) we identify target APIs by examining browser behavior during manual website interaction.

Having identified a target API, EDEFuzz's workflow consists of four main steps which are in turn divided into two phases: recording/preparation phase (Step 0) and relaying/fuzzing phase (Steps 1-3), in line with its record/replay design discussed in **Section 1**.

**Recording/Preparation Phase.** In this semi-automated phase, the goal is to generate a configuration file denoted as $C$ that brings the client under test to a baseline state. This file will be used in the subsequent replaying/fuzzing phase. To that end, we use a Web Proxy to capture the traffic between a client app, which is a web browser in our experiments, and the targeted web server. Specifically, we start the client app and capture its initial state $S_0$. After that, we let the client open the given URL, wait for the web page to be fully loaded. The tester then interacts with the page (e.g., fill in text boxes, click buttons) to trigger a request to the target API. We denote $S_1$ as the client state at which the request has just been completely sent. We develop a lightweight browser plugin to capture the interaction steps required to traverse from state $S_0$ to $S_1$ and save them into the configuration file so that they can be played back in subsequent steps.

The standard most common for responses to web API requests is JSON. Under AJAX or similar paradigms, when the API response is received a client uses the JSON response to update the web page (e.g., showing more information) and the update typically alters only some parts of the web page, rather than changing the entire page. We denote $S_2$ as the client state immediately following the update. This state can be typically identified by the existence of certain page elements. The steps required to identify the transition from $S_1$ to $S_2$ (typically of the form "wait until page element $X$ appears") are also recorded in the configuration file, meaning it now stores all steps required to traverse from state $S_0$ to $S_2$. At $S_2$, a baseline DOM tree (denoted as $\mathcal{D}_{\text{origin}}$) of the web page is extracted using the DOM Extractor component. This DOM tree will be compared with other trees to be generated in the fuzzing phase to check for potential excessive data exposures based on the metamorphic relation defined in **Equation (1)**.

Before the targeted request-response pair has been exchanged, the browser and the server might have completed other exchanges for fetching HTML documents and other resources such as images, stylesheets and Javascripts. Thus all request-response pairs and resources, denoted $P$ are recorded and stored for the replaying/fuzzing phase.

**Replaying/Fuzzing Phase.** The input for this phase includes: 1) the configuration file $C$, 2) the original DOM tree $\mathcal{D}_{\text{origin}}$, and 3) all request-response pairs $P$ recorded in the recording phase.

**Listing 2: A configuration file generated in recording phase.**

```
1  TARGET /api/v2/stock/get
2
3  LOAD https://www.example.com/path/page
4  INPUT //input[@id="text-postcode"] 3000
5  CLICK //span[text()="Check availability"]
6  WAIT_LOCATE //div[@id="stock-info"]/div[2]
7  FUZZ
```

**Listing 2** shows a sample configuration file. The first line specifies the target API under test. The rest define a sequence of user interactions to be performed to reach state $S_2$. The example loads a web page, enters the number 3000 into a text box, clicks a button and then waits for a specific element to appear on the web page before capturing the state. Apart from these actions, EDEFuzz also supports HOVER, SLEEP, and SCROLL for (i) hovering the mouse on a specific element, (ii) waiting for a specified period of time, and (iii) scrolling up/down, respectively.

In each fuzzing iteration, EDEFuzz goes through three steps (Steps 1-3). In the first step (Step-1), the Web Driver component, which is built on top of the Selenium Web Driver [18], uses the configuration file $C$ to replay all the steps until the client reaches the state $S_1$. Before that state has been reached, the Simulated Server responds to client requests with the corresponding recorded responses stored in $P$ with no modification. Once state $S_1$ is reached and the Simulated Server receives the request sent to the targeted API, it mutates the originally recorded response by deleting a specific data field and transmits to the client. We describe the mutation algorithm in detail in **Section 4.3**.

After the baseline state is reached, in Step-2 the client uses the mutated response to update the page accordingly. If this leads to any error, EDEFuzz moves to the next fuzzing iteration. Otherwise, EDEFuzz waits until the page is fully updated (i.e. until state $S_2$ is reached) and uses the DOM Extractor to extract the current DOM tree denoted as $\mathcal{D}_{\text{mutated}}$. In Step-3, EDEFuzz compares $\mathcal{D}_{\text{mutated}}$ and $\mathcal{D}_{\text{origin}}$ using a comparison algorithm described in **Section 4.4**. If the two DOMs are the same (based on our definition of similarity) then we flag the deleted data field. According to the metamorphic relation, the field is excessive. Once all fuzzing iterations are completed, EDEFuzz reports all the potential excessive data fields to the tester for further analysis and confirmation (see **Section 4.5**).

*Randomness:* The explanation so far assumes that web replaying is fully deterministic: given a configuration file, EDEFuzz—without applying any mutations on the server response—produces the same DOM tree across all runs. Though uncommon, we identified a few cases in which this assumption does not hold. For instance, a social media platform may randomly insert advertisement between user posts during runtime. This causes the web page to be visually different in each run, even if all contents transmitted from the server were identical. We discuss the issues in details and how we address them in Section 4.2 and Section 4.4.

**Implementation Details.** We implemented EDEFuzz in Python3 using Selenium as a Web Driver to control web browser. We successfully tested our tool on widely used platforms including Ubuntu 18.04, Ubuntu 20.04 and Windows 10. Currently, EDEFuzz supports two web browsers: Google Chrome and Mozilla Firefox. Our design is modular to support future extensions.

## 4.2 Simulated Server

Our design using a Simulated Server brings several benefits to EDE testing. The server supplies locally recorded contents with minimal delay—reducing test latency. Secondly, EDEFuzz does not need to communicate with the targeted remote server during the fuzzing phase, allowing testing in parallel without affecting the targeted server. This allows developers to easily test their servers without impacting production services. This design also ensures consistency within a given run, ensuring test results are not affected by potential changes to the state of the remote server during the testing process.

Since the simulated server provides a snapshot of states on the remote server, there is no need to handle cookies or sessions within EDEFuzz. The simulated server matches the requested URL from the collection of recorded request-response pairs and simply sends back the recorded response. This also supports sites the user needs to first log-in: as long as the remote server sent responses representing the session of a user who was logged-in during the recording phase, our simulated server can reproduce it in the fuzzing phase.

Websites that fetch resources by randomised URLs instead require a fuzzy matching algorithm in the simulated server. A typical case is when requesting a resource from the server, the web page generates a random token to be included in the request URL. This case causes a request to be generated for which the simulated server has no recorded response. Our experimental evaluation in Section 5.2 shows that this applies to only a fraction of popular websites (8.7% of our evaluated target set). We believe it can be handled by applying fuzzy matching for request URLs, but leave this extension for future work.

## 4.3 Mutation Engine

The original API response (JSON) produced by the server is assumed to be valid both structurally and semantically. EDEFuzz generates test cases by mutating the original API response.

We represent the server-supplied JSON object using a tree. Each leaf node is potentially an excessive data field. For example, Figure 4 shows the (partial) tree representation of the JSON object from Company-I shown in Listing 1. In this tree, all leaf nodes are shaded.

We generate each mutation (test case) by removing a leaf node from the tree. For example, a valid mutation of the JSON object from Company-I shown in Listing 1 could remove the key-value pair id: 353 from the driver dictionary, or capacity: 33 from the car dictionary.

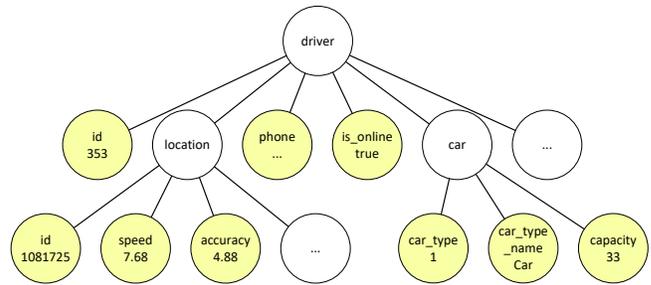

Figure 4: Tree representation of the JSON object in Listing 1. The mutation engine uses this represenation to determine what to mutate. We highlight leaf nodes in yellow.

Unlike other fuzzing approaches which may generate an infinite number of test cases (e.g by using genetic mutation operators such as bit flips and splicing [19]), our approach produces a fixed number of test cases based on the actual leaf nodes on the tree. Standard techniques to further reduce that number using approaches like binary tree search and delta debugging [20] are not applicable because there is no way to determine whether a subtree (or set of fields) contains at least *one* field that constitutes an EDE (if we delete all of them and notice a change in the DOM, then it might mean that none are excessive, or all-but-one are excessive, or something in-between). Instead our metamorphic relation can tell us only if *all* fields in some set are subject to EDE (when we remove them all and no change is detected in the DOM). It is worth noting that this design decision also ensure more predictable test run times.

## 4.4 Similarity Check for DOM Trees

A web page (an HTML document) has a hierarchical structure that can be represented using a DOM tree. In Section 4.4 and Figure 5 we show a sample HTML snippet and its DOM tree, respectively. This is a simplified version of one of our testing targets Company-I, which we shared in Section 3.

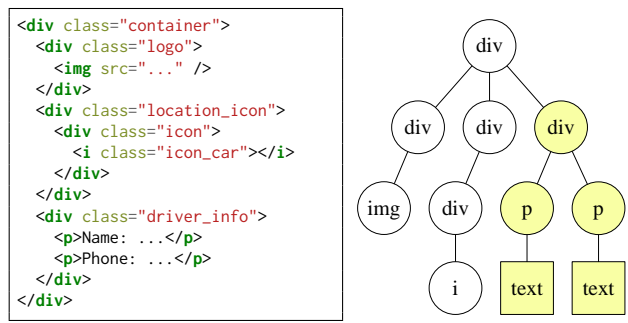

Figure 5: The tree representation of a simple HTML snippet. Each rounded node represents a tag element while each square node represents a string. The hierarchical structure allows a human to easily specify a subtree (in yellow) for the tool to evaluate.

Recall, a DOM tree has nodes and each tree node represents a tag element in the HTML document. Each tree node has zero or more attributes—which are stored as key-value pairs—and each node may have children, which are either further tree nodes or plain strings.

When comparing two web pages, one considers both the DOM tree structure and the content within each tree node. According to our metamorphic relation, if an API response with a particular data field removed would still result in the identical DOM tree, compared to the DOM tree produced using the original API response, that data field is reported as excessive. Two DOM trees are considered identical if and only if *all* of the following conditions hold:

- **(C1)** Their root nodes have the same tag name
- **(C2)** Their root nodes have the same number of attributes
- **(C3)** Each corresponding pair of attributes have the same value
- **(C4)** Their root nodes have the same number of children
- **(C5)** Each corresponding pair of children representing a tag element is identical, with respect to conditions **C1-4**
- **(C6)** Each corresponding pair of children representing a string is identical

To check for all of these conditions, the most simple yet effective approach is to compare the string representations of the corresponding HTML documents. Our experiments showed that this works for 47 out of 54 targets. However, we observed that several web pages contain elements which are not affected by the API response. For example, the value of the `class` attribute within a `<div>` tag could be randomly generated at run-time. Another common case is when the web page displays the current date and time on it. Apparently, these cases could yield false negatives using the straightforward string-based comparison approach.

To address this issue, we relax the conditions **C3** and **C6**. That is, we accept the differences in string leaf nodes and in attributes' values caused by randomness. To that end, EDEFuzz runs a pre-processing step before the fuzzing phase. In this step, EDEFuzz uses the configuration file *C* to replay and generate a few DOM trees, all generated from replaying the *same* server response. After that, it recursively traverses and compares $\mathcal{D}_{\text{origin}}$ with each of the newly generated DOM trees to look for those parts that differ due to randomness, and marks those random elements and attributes of $\mathcal{D}_{\text{origin}}$ that should be "ignored" in the comparison step of the fuzzing phase. It is worth noting that, in this pre-processing step, if EDEFuzz finds that the generated DOM trees are *structurally* different from $\mathcal{D}_{\text{origin}}$, i.e. violating conditions **C1**, **C2**, **C4** and **C5**, it will decide to terminate the testing process. In our experiment, 9 of 69 targets were flagged at this step, meaning that they each produced web pages with different structures, even using the same response.

In the fuzzing phase, EDEFuzz compares the DOM tree $\mathcal{D}_{\text{mutated}}$ produced from each test case, with the pre-processed $\mathcal{D}_{\text{origin}}$. Basically, EDEFuzz (i) recursively traverses through every node in $\mathcal{D}_{\text{origin}}$ and its corresponding node in $\mathcal{D}_{\text{mutated}}$, and (ii) compares each pair of nodes. EDEFuzz will skip all string nodes and attributes on $\mathcal{D}_{\text{origin}}$ that have been marked as "ignored" in the pre-processing step. Two nodes are deemed structurally different if any of the conditions **C1**, **C2**, and **C4** is violated. Moreover, they are considered different in terms of content if there is a discrepancy in the values of the nodes (in the case of non-ignored string nodes) or the non-ignored attributes (for other types of nodes).

While comparing the entire DOM tree can identify if a web page is different from another, in many cases a response will affect only specific areas of a web page. Our approach can optionally utilise human knowledge to allow the user to specify an *area-of-interest* on the web page. The area-of-interest is a subtree in the DOM tree that contains contents (that the user believes are) affected by the API response. The area-of-interest in the **Figure 5** example could be the subtree rooted at the node `<div class="driver_info">` (highlighted in yellow in the tree representation). This helps increase efficiency, and it also avoids other components on the web page affecting the comparison (e.g. if they are randomly generated).

### 4.5 Result Inspection

The final step of our approach is manually inspecting the results. This involves inspecting each of the flagged data fields to determine what kind of data it exposes (e.g. sensitivity) and, therefore, whether the web application should be modified to avoid this exposure.

We suggest the following approach to determine whether a given field is indeed unnecessary (i.e., a true positive), and how sensitive the data is (the severity of a leak).

**False positives.** Like other testing/fuzzing approaches, EDEFuzz could yield false positives. For instance, a value included in an API response may only be populated on the web page with further user interactions. This is not uncommon when using an overlay window to display certain information. A tester typically verifies a flag by further interacting with the web-page, discovering any hidden dependencies on the field.

**Sensitivity of the excessive data.** To determine the sensitivity of an excessive data field, first we can assess whether the data field is human interpretable. This may include a minimal decoding process to parse information represented under a given scheme (eg; base64, GPS coordinates, Unix timestamp, gzip). Next, we determine whether the user could legitimately acquire the data some other way. If not, we consider it 'sensitive'. We present real examples from our dataset. We note that some non-interpretable may nonetheless be sensitive when parsed appropriately. We merely present a usable heuristic applicable to our evaluation.

## 5 EXPERIMENTAL EVALUATION

We designed our experimental evaluation to answer the following research questions.

**(RQ1) Accuracy.** Of the data fields flagged by EDEFuzz, what proportion are true excessive data exposures (i.e. unused by the web page). This evaluates the usefulness of our metamorphic relation.

**(RQ2) Applicability.** To what proportion of widely used web sites can EDEFuzz be applied successfully? This helps to understand limitations of our approach, both inherent and those that arise from EDEFuzz's current implementation.

**(RQ3) Efficiency.** How much human effort and computational time is required to apply EDEFuzz? This sheds light on the scalability of our approach.

**(RQ4) Prevalence of Sensitive Data Leakage.** Of those fields flagged by EDEFuzz as excessive, what proportion contain sensitive data? This helps us understand how prevalent sensitive data leakage is amongst excessive data exposure issues.

Note the distinction between **RQ1** and **RQ4**. Specifically, given some set of data fields reported by EDEFuzz as excessive, the former measures the proportion of *true positives* whereas the latter measures the proportion that are *sensitive*. These notions are entirely orthogonal: a flagged data field is a true positive precisely when it is really excessive (i.e. unused by the web page), regardless of whether the data it contains is sensitive or not. Likewise, a field contains sensitive data precisely when that data should not be revealed by the web application, regardless of whether the field is excessive (i.e. is unused by the web page) or not. Flagged fields that are not true positives are *false positives*. False positives can exist, for instance, when a field contained in a response is used to affect the DOM only after subsequent user interaction with the web page. Importantly, whether a data field is a true or false positive is a property of the behaviour of the web page. That is, distinguishing between true and false positives requires carefully understanding *all* behaviours of the web application, including by reading its code and interacting with it, to determine whether the data field contained in the response is ever used in future by the page. Whether a field is sensitive or not is simply a property of the data that field contains, and can be relatively quickly ascertained by inspecting just the field itself.

| Target    | Rank | Used when                         |
|-----------|------|-----------------------------------|
| Company-A | 37   | Load tracking history of parcel   |
| Company-B | –    | List members of a school subject  |
| Company-C | 121  | Check stock availability of item  |
| Company-D | –    | Check stock availability of item  |
| Company-E | 59   | Check stock availability of item  |
| Company-F | 166  | Get flight prices for next 30 days|
| Company-G | 123  | Check stock availability of item  |
| Company-H | 2770 | List vehicles available for sale  |

**Table 1: Australian websites tested. We include short descriptions of the purpose of the targeted APIs on which we performed deeper evaluation. Where available we provide Alexa rankings within Australia (extracted 29th April 2022).**

This means that accurately evaluating **RQ1** requires a set of web sites whose behaviour is (or can be) well-understood by the humans inspecting EDEFuzz's results. Carefully understanding the behaviour of an individual site can be very time consuming. Therefore, for **RQ1** we assembled a set (see Table 1) of eight (8) popular websites within a single country (Australia) that were familiar and whose individual behaviours of the web page were therefore able to be well-understood. This purposeful restriction was necessary to ensure that the proportion of true positives could be accurately evaluated.

In contrast, adequately evaluating the remaining research questions requires a data set that comprises widely used web sites. Therefore the remaining RQs were evaluated on a data set drawn from the Alexa Top-200 list of web sites. While not applicable to **RQ1** (since enumerating all of their possible behaviours to accurately determine true positives is infeasible), this set forms a representative best-of sample, suitable for assessing EDEFuzz's performance in general (**RQ2** and **RQ3**), as well as to understand the prevalence of sensitive data leakage via EDE among common web sites (**RQ4**).

### 5.1 Procedure

We follow the 4-step procedure below to fuzz test a given website using EDEFuzz. The procedure is well-aligned with the workflow of the tool, as discussed in Section 4.1.

**(Step-1) Identifying the target API endpoint.** Given a primary domain, we identify API endpoint(s) for inclusion in the testing regime. Each domain accessed a variety of APIs, some internal to the domain (eg., a shopping website accessing the site's stock inventory and pricing endpoint) and some external (eg., an analytics endpoint to retrieve recent visitor counts). Of these, we manually identify the internal API most relevant to the function of the site in question. For instance, in Table 1 we list the selected Australian sites and endpoints used in our evaluation.

**(Step-2) Writing a configuration file.** We compose a configuration file that specifies how to correctly trigger the selected endpoint—this includes the sequences of user interactions that preclude the execution of a request to the API.

This process is semi-automated via our custom-built web browser plugin. However the plugin still requires the user to manually perform interactions. After the file is generated, the operator reviews the file to make any necessary changes.

We specified an area-of-interest (see Section 4.4) on all our evaluation targets, at the cost of a few seconds per target.

**(Step-3) Running EDEFuzz.** After devising an appropriate configuration, we then execute EDEFuzz, time its execution, and collate the results for analysis. We chose not to repeat the fuzzing process for each target because, unlike in traditional fuzzing, EDEFuzz's mutation process is deterministic by design.

**(Step-4) Analysing results.** This involves the manual classification step, discussed in Section 4.5.

We adhere to a systematic classification model to decide if a reported EDE is sensitive or not. We first assess whether the data field is readily human interpretable (including simple encoding schemes such as base64, timestamps). We mark non-human interpretable values non-sensitive. Next, we determine whether the user could legitimately acquire the data otherwise. If yes, mark the field non-sensitive; if not, mark it sensitive.

### 5.2 Results

*RQ1. Accuracy.* We evaluated the accuracy of our metamorphic relation, as implemented in EDEFuzz, for identifying excessive data exposures against the eight sites listed in Table 1. The TP column of Table 2 summarises the results by recording the true positive rate, namely the proportion of reported excessive fields (Reported) that were actually excessive (Confirmed). This was determined by manually inspecting the web pages and carefully understanding their behaviours, including via manual interaction, to check whether, for each reported data field, they had any behaviours that made use of the data field. If no such behaviours were identified, the field was classified as a true positive; otherwise it constitutes a false positive.

The overall true positive rate was 98.65%, confirming the exceptionally high degree of accuracy of our approach. (Later, **RQ4** investigates which of these EDEs actually leaked sensitive data which, as argued in Section 5, is a separate concern.)

EDEFuzz, like other testing/fuzzing approaches, may yield false positives. For instance, a value included in an API response may only be populated on the web page after further user interaction. This pattern arises for instance when using an overlay window to display certain information. We manually investigated each false positive (1.35% overall) and found that each was an instance of this pattern.

*RQ2. Applicability.* We evaluated the applicability of EDEFuzz on a subset of the top 200 sites (as recorded by the Alexa ranking). Of the 200 sites, we excluded (see Figure 6) from analysis those that had no web APIs (and, hence, no possibility for EDE); those requiring payment; those in a language that none of our authors understand; those comprising adult or illegal content; those that were geoblocked; and those that required solving a CAPTCHA. Doing so excluded around 60% of the 200 sites, after deduplication. None of these exclusions represent limitations of our approach or implementation. Of the remaining sites, we additionally excluded 12 sites that used HTTP_POST requests to query their APIs with query parameters included in the request body, since EDEFuzz's simulated server currently relies solely on the request URL to supply the response, though this implementation-level limitation could be resolved with modest future work. This left 69 sites for our evaluation of EDEFuzz.

| Target | Data fields | Reported | Confirmed | TP | Preparation (min) | Execution (min) | Classification (min) | Sensitive | Non-sensitive |
|---|---|---|---|---|---|---|---|---|---|
| COMPANY-A | 189 | 124 | 124 | 100.00% | 10 | 11 | 5 | 0 | 124 |
| COMPANY-B | 18 | 16 | 14 | 87.50% | 20 | 2 | 2 | 2 | 12 |
| COMPANY-C | 2600 | 2580 | 2504 | 97.05% | 5 | 306 | 3 | 104 | 2400 |
| COMPANY-D | 545 | 506 | 479 | 94.66% | 15 | 43 | 10 | 9 | 470 |
| COMPANY-E | 4249 | 4147 | 4127 | 99.52% | 10 | 755 | 15 | 0 | 4127 |
| COMPANY-F | 778 | 749 | 749 | 100.00% | 15 | 103 | 5 | 0 | 749 |
| COMPANY-G | 120 | 100 | 100 | 100.00% | 5 | 12 | 3 | 0 | 100 |
| COMPANY-H | 1465 | 1066 | 1066 | 100.00% | 15 | 79 | 20 | 19 | 1047 |

Table 2: Summary statistics from the Australian sites. Data fields reports the total number of fields contained in the API response of each target, Reported is the number of fields flagged by EDEFuzz as excessive; Confirmed is the number of fields manually confirmed to be excessive, i.e. true positives, TP. The time taken to configure EDEFuzz for each target is reported in Preparation, as is the duration of test execution (Duration) and the human effort required to manually classify the flagged fields as sensitive or not (Classification), all measured in minutes. We also report (Sensitive) the number of fields we classified as containing sensitive data, after manual inspection.

Of these 69 sites, EDEFuzz successfully applied to 53 targets (76.8%). Of the 16 unsuccessful targets, all but one were the result of nondeterminism aka randomness: nine (13.0%) did not pass the pre-processing step (as explained in Section 4.4), as these websites populated elements of their page non-deterministically; six (8.7%) used requests that included non-deterministic tokens required to load resources. The final unsuccessful target used shadow DOM within its web page, preventing EDEFuzz from accessing its complete DOM tree.

We further assessed applicability by performing an additional validation step on those 53 successful targets. This was done to test the implicit assumption of our mutation strategy that, given any two fields of a response, whether one is excessive is independent of the presence or absence of the other, i.e. that each field of a response can be assessed independently of the others. We tested this assumption for each of the 53 successful targets by taking the entire set of fields flagged by EDEFuzz as excessive and removing *all* of them simultaneously from the recorded response, and replaying that mutated response to see whether the resulting DOM passed the similarity check (Section 4.4) to the original. Surprisingly we found that for three (3) of the sites, removing multiple fields at once caused a difference in the DOM even when removing each of the fields individually did not (behaviour we confirmed via manual testing for each of them). For these sites, one might reasonably debate whether those flagged fields really are excessive or not, a question we leave for future work.

EDEFuzz is highly applicable; however sites that employ randomness, either in their DOM *structure* or the requests they generate, are beyond EDEFuzz's current design. The latter we posit could be addressed by fuzzy matching in the simulated server (see Section 4.2).

*RQ3. Efficiency.* Efficiency measures not only the amount of computational time to employ EDEFuzz, but also the amount of human effort both to configure the tool and to inspect its results to determine which reported excessive data fields contain sensitive data. We evaluated efficiency across both the Australian sites (Table 2) and the Top 200 data set. Experiments for the Australian sites were carried out on a commodity PC with an Intel Core i7-9600K, 32GB of RAM, running Windows 10 21H1. Those for the Top 200 data set were carried out on an AWS VPS with a 16-core Intel CPU and 32 GB of memory, running Ubuntu 20.04.

The time spent on test execution was roughly linear in the number of data fields included in the response from the target API, as expected since our mutation strategy must necessarily mutate one field at a time. On average, about 8 test cases were executed per minute and this figure is consistent between the two data sets. However, computation time does not present a significant bottleneck, especially since our approach is trivially parallelised.

Regarding human effort, it took a maximum of twenty minutes per web site for us to identify an appropriate API endpoint that it made use of and to then compose a configuration file, representing minimal overhead.

While naturally some human effort is required to inspect the flagged fields, to determine which are sensitive, this again took no longer than 20 minutes per web site, even when EDEFuzz reported many thousands of fields as excessive. This was because many API responses contained large numbers of repeated structures, which allowed us to quickly classify thousands of flagged fields. However, we found certain flagged data fields required lengthier and more comprehensive analysis. We hypothesise that this is a fundamental limitation on automation (for the near future) as deciding whether a field is indeed sensitive is an exercise of human judgement, involving considering the application's function, what data is already publicly available, and privacy expectations, etc.

Overall we conclude therefore that our approach requires only a modest amount of human effort.

*RQ4. Prevalence of Sensitive Data in EDEs.* Finally, our results allow us to draw conclusions about the prevalence of sensitive data leakage via excessive data exposure. Such conclusions necessarily *underestimate* the true extent of sensitive data leakage, even on the sites used in our evaluation, since we tested on each site only one—highly-visible and, hence, likely to be widely-tested—API.

Among the Australian websites, we find that sensitive data leakage is much more prevalent (present in 4 out of the 8 cases evaluated) than among the Alexa Top 200 sites (where it is present in only one of the 53 cases successfully evaluated). We conjecture that this should be expected, since popular sites are more widely used (and thus tested) by definition. Yet even among very popular sites, EDEFuzz still found sensitive data leakage.

We already discussed sensitive data leakage discovered by EDE-Fuzz in the Australian websites Company-C, Company-D, and Company-E in Section 3. (We also discussed vulnerabilities it found in Company-I and Company-J during testing, which are not part of our evaluation set). EDEFuzz also identified sensitive data leakage in Company-B, the learning management system with the largest market share ($\approx$ 34%). Company-B has a feature to create student groups within a subject. Instructors of a subject can view and assign group members of each group. Our tool flagged the API that lists group members. While the web page only displayed a list of names in a group, the API response contained the full list of subjects each student is enrolled into. We further found that this API is accessible from a student account as well, allowing any student to gain knowledge about into which subjects their classmates have ever enrolled.

The one instance of sensitive data leakage found by EDEFuzz in the Alexa Top 200 (Rank 91) affected an API called by a web page for downloading device drivers, which inadvertently exposed employee names as excessive data, as well as hardware IDs.

***Summary.*** EDEs appear prevalent, though many are relatively harmless. However, much like memory corruption vulnerabilities (whose severity can vary wildly), they can also be severe and leak sensitive information. EDEFuzz is effective at diagnosing such vulnerabilities via its highly accurate metamorphic relation, requiring modest human effort and computational cost, while being widely applicable.

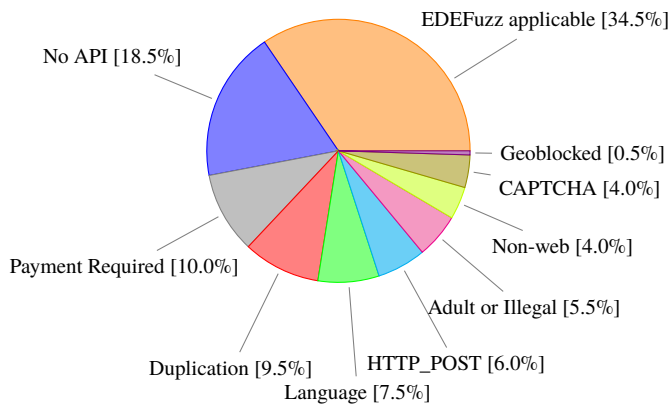

**Figure 6: Applicable websites from Alexa Top-200. Of the 200 we found 69 (34.5%) appropriate for our testing-set. We excluded domains for the following reasons: duplication of a single service across domains, adult and illegal content, geoblocking, payment required for access, lack of an API, foreign language and encoding of parameters in POST requests (discussed in Section 5.2).**

## 6 RELATED WORK

**Detecting Vulnerabilities in Web Applications** Researchers have made satisfying progress in detecting and preventing *certain classes* of webapp vulnerabilities. Much work [21]–[25] exists for detecting cross-site scripting [26]. The techniques vary, including black-box [22], [25], grey-box [23] and white-box [21] approaches. Very recently, Trickel, et al. [27] proposed a novel grey-box fuzzing approach to detecting SQL and command injections.

However, excessive data exposure has received comparatively little attention despite being one of the most common vulnerabilities [28]. To the best our knowledge, there is no published research focusing on an automated approach to detect this class of vulnerability in web applications: our tool EDEFuzz is the first of its kind.

Koch, et al. [29] studied a white-box and semi-automated mechanism to identify EDE vulnerabilities in Android applications—not in web applications. Unlike EDEFuzz which runs the program on two different outputs from the web server (one with a data field deleted) and looks for the absence of difference in the DOM to detect leakage, their approach requires (decompiled) source code of the applications to do instrumentation and static data-flow analysis. Its static analysis identifies potential EDEs by flagging data received by the app over the network (source) that is then serialised to a Java object but that then never propagates to the user interface (sink). A subsequent dynamic analysis that relies on program instrumentation and manual app interaction is used to confirm potential vulnerabilities, wherein the human analyst must manually generate tests that attempt to trigger the EDE. EDEFuzz also requires manual effort to interact with a web application to trigger the web API under test, and like Koch et al. also to confirm the sensitivity of leaked information. So [29] and our approach are complementary. As demonstrated in fuzzing research [13], combinations of complementary approaches could yield better results and we leave that for future work.

**Metamorphic Testing/Fuzzing** As discussed in Section 2, one of the most important steps of metamorphic testing is to identify metamorphic relation(s). This requires creativity and a good understanding of the system under test. To ease this crucial step, Segura, et al. [30] proposed six abstract relations from which concrete relations can be defined. Specifically, the authors identified 60 API-specific metamorphic relations in their work. Their relations specify how related web requests should produce related responses, and so are inapplicable to detecting EDE. The fundamental insight of EDEFuzz that shows how metamorphic fuzzing is applicable to detecting EDE is not to mutate the *request* (as in all prior web fuzzing work), but to mutate the *response* instead.

**RESTful Web API Testing** RESTler [6]—the state-of-the-art RESTful API fuzzing approach—used server states, relying on response codes to identify server crashes on APIs used by cloud services. Their tool infers dependencies among requested APIs to guide the generation of new test cases. Atlidakis, et al. [11] suggested an extension to RESTler to report the violation of four rules commonly applied to REST services, in addition to server crashes. [31] improved RESTler's test generation algorithm by representing the payloads in a tree structure on which structure-aware mutation operators can be applied. Pythia [10] augmented RESTler with coverage-guided feedback and it implemented a learning-based mutation strategy. Specifically, it utilised a statistical model to gain knowledge about frequent ordering of calling APIs from seed inputs. It used a regular grammar to encode an API request and perform mutation by adding noise to the request. Neither RESTler nor its follow-up works can detect EDE vulnerabilities. Moreover, this line of work focuses on mutating the API requests while EDEFuzz modifies the API responses.

**Record-replay Mechanism in Web Application Testing** Record-replay models are popular in testing web applications and services. We use two types of record-replay in our work. The first bears similarity to WaRR [32], which records the interaction between a user and a web application. It allows the recorded traces to be later replayed to simulate the user interacting with the web application. Another similar work is Timelapse [33], in which researchers log unexpected behaviours in web applications to help developers visualise, demonstrate and better understand bugs. We follow their mechanism, automating interactions with a web application through a headless web-driver. The second form of record-replay tools capture communications between a server and client, to be replayed at a later time [7]. While existing work focuses on producing an exact replication during a replay stage, EDEFuzz uses a simulated server to instead supply mutated server responses.

**Web Change Detection** The components of a web application may change over time, hindering research and testing that relies on consistency. Researchers have looked into different strategies to compare two pages and identify their differences. One strategy was proposed over twenty years ago and relies on the HTML DOM tree to monitor structural changes on the web page [34]. Other relevant work includes X-Diff [35] which aimed at detecting changes in an XML document, and [36] which improved the efficiency of Hungarian algorithm in detecting web page changes. Modern web applications have increased in complexity and often break from prior design paradigms. As a result, past approaches for page comparison are less effective than they once were. To address this challenge, Waterfall [37] uses two versions of the same web application, in detecting locator changes and applying fixes. WebEvo [38] attempts to identify evolution of a web application through detecting semantic structure changes. Both approaches aim at matching contents between two structural different web pages, while our work focus on identifying both structural difference and content difference.

## 7 DISCUSSION AND FUTURE WORK

Simple ideas are often the best. From our evaluation we conclude that EDEs appear prevalent, and that EDEFuzz is effective at finding them, including sensitive data leakage, with acceptable efficiency and requiring a modest amount of human work. Its metamorphic relation yields precise results in practice (a TP rate of 98.65%). It is also generally applicable, and it can be parallelized easily. At the same time, our results suggest avenues for improvement.

**Handling HTTP POST Requests** by parsing request bodies, which affected 6% of the targets from the Alexa Top-200 websites.

**Improving Efficiency.** Even though our Simulated Server helps EDEFuzz achieve a reasonable fuzzing throughput, we can improve it further by leveraging the recent advancement in the topic of snapshot-based fuzzing [39]. As the design of EDEFuzz is modular, the change could be minimal. Specifically, we can take a snapshot of the client at the state $S_1$ when a request to the target API has just departed (See Section 4). We can then restore the snapshot for each fuzzing iteration instead of replaying requests using the Web Driver.

**Fuzzy Matching.** We posited in Section 4.2 that EDEFuzz's current Simulated Server has implementation limitations that prevent it from being able to respond to randomised requests. It could be overcome by implementing fuzzy matching. While affecting only 8.7% of our sites, this is a straightforward avenue for future investigation.

## ACKNOWLEDGMENT

We thank the anonymous reviewers for their constructive and encouraging feedback. We also thank Amazon (through their Research Program) for giving us AWS Cloud Credit to conduct our large-scale experiments.

## REFERENCES

[1] *OWASP API Top 10: 2019*, 2019. [Online]. Available: https://owasp.org/www-project-api-security/ (cit. on p. 1).

[2] *Microsoft announces new Project OneFuzz framework*, 2020. [Online]. Available: https://www.microsoft.com/security/blog/2020/09/15/microsoft-onefuzz-framework-open-source-developer-tool-fix-bugs/ (cit. on p. 1).

[3] E. T. Barr, M. Harman, P. McMinn, M. Shahbaz, and S. Yoo, The oracle problem in software testing: A survey, *IEEE transactions on software engineering*, vol. 41, no. 5, 507–525, 2014 (cit. on p. 1).

[4] T. Y. Chen, S. C. Cheung, and S. M. Yiu "Metamorphic testing: A new approach for generating next test cases," Tech. Rep. HKUST-CS98-01, 1998, arXiv preprint arXiv:2002.12543 (cit. on pp. 1, 3).

[5] *Using burp to test for sensitive data exposure issues*, 2022. [Online]. Available: https://portswigger.net/support/using-burp-to-test-for-sensitive-data-exposure-issues (cit. on pp. 2, 4).

[6] V. Atlidakis, P. Godefroid, and M. Polishchuk Restler: Stateful rest api fuzzing, in *2019 IEEE/ACM 41st International Conference on Software Engineering (ICSE)*, IEEE, 2019, 748–758 (cit. on pp. 2, 3, 10).

[7] R. Netravali, A. Sivaraman, S. Das, et al. Mahimahi: Accurate {record-and-replay} for {http}, in *2015 USENIX Annual Technical Conference (USENIX ATC 15)*, 2015, 417–429 (cit. on pp. 2, 11).

[8] S. Ikeda, Massive optus data leak prompts new privacy rules in australia, *CPO Magazine*, 2022 (cit. on p. 2).

[9] *Api3:2019 excessive data exposure*, 2019. [Online]. Available: https://github.com/OWASP/API-Security/blob/master/2019/en/src/0xa3-excessive-data-exposure.md (cit. on p. 3).

[10] V. Atlidakis, R. Geambasu, P. Godefroid, M. Polishchuk, and B. Ray, Pythia: Grammar-based fuzzing of rest apis with coverage-guided feedback and learning-based mutations, *arXiv preprint arXiv:2005.11498*, 2020 (cit. on pp. 3, 10).

[11] V. Atlidakis, P. Godefroid, and M. Polishchuk Checking security properties of cloud service rest apis, in *2020 IEEE 13th International Conference on Software Testing, Validation and Verification (ICST)*, IEEE, 2020, 387–397 (cit. on pp. 3, 10).

[12] OpenAPI Initiative Openapi specification, Available online: https://swagger.io/resources/open-api/, 2020 (cit. on p. 3).

[13] V. J. M. Manès, H. Han, C. Han, et al., The art, science, and engineering of fuzzing: A survey, *TSE*, 2019 (cit. on pp. 3, 10).

[14] *Google clusterfuzz*, 2022. [Online]. Available: https://google.github.io/clusterfuzz/ (cit. on p. 3).

[15] P. He, C. Meister, and Z. Su Testing machine translation via referential transparency, in *2021 IEEE/ACM 43rd International Conference on Software Engineering (ICSE)*, IEEE, 2021, 410–422 (cit. on p. 3).

[16] M. Rigger, and Z. Su Testing database engines via pivoted query synthesis, in *14th USENIX Symposium on Operating Systems Design and Implementation (OSDI 20)*, 2020, 667–682 (cit. on p. 3).

[17] *What is social engineering?* 2022. [Online]. Available: https://www.kaspersky.com.au/resource-center/definitions/what-is-social-engineering (cit. on p. 4).

[18] *Webdriver*, 2022. [Online]. Available: https://www.selenium.dev/documentation/webdriver/ (cit. on p. 5).

[19] *American fuzzy lop*. [Online]. Available: https://lcamtuf.coredump.cx/afl/ (cit. on p. 6).

[20] A. Zeller, Yesterday, my program worked. today, it does not. why? *ACM SIGSOFT Software engineering notes*, vol. 24, no. 6, 253–267, 1999 (cit. on p. 6).

[21] R. Wang, G. Xu, X. Zeng, X. Li, and Z. Feng, Tt-xss: A novel taint tracking based dynamic detection framework for dom cross-site scripting, *Journal of Parallel and Distributed Computing*, vol. 118, 100–106, 2018 (cit. on p. 10).

[22] F. Duchene, S. Rawat, J.-L. Richier, and R. Groz Kameleonfuzz: Evolutionary fuzzing for black-box xss detection, in *Proceedings of the 4th ACM conference on Data and application security and privacy*, 2014, 37–48 (cit. on p. 10).

[23] O. v. Rooij, M. A. Charalambous, D. Kaizer, M. Papaevripides, and E. Athanasopoulos Webfuzz: Grey-box fuzzing for web applications, in *European Symposium on Research in Computer Security*, Springer, 2021, 152–172 (cit. on p. 10).

[24] U. Sarmah, D. Bhattacharyya, and J. K. Kalita, A survey of detection methods for xss attacks, *Journal of Network and Computer Applications*, vol. 118, 113–143, 2018 (cit. on p. 10).

[25] A. Doupé, L. Cavedon, C. Kruegel, and G. Vigna Enemy of the state: A {state-aware} {black-box} web vulnerability scanner, in *21st USENIX Security Symposium (USENIX Security 12)*, 2012, 523–538 (cit. on p. 10).

[26] M. C. Martin, and M. S. Lam Automatic generation of xss and sql injection attacks with goal-directed model checking. In *USENIX Security symposium*, 2008, 31–44 (cit. on p. 10).

[27] E. Trickel, F. Pagani, C. Zhu, ET AL. Toss a fault to your witcher: Applying grey-box coverage-guided mutational fuzzing to detect sql and command injection vulnerabilities, in *2023 IEEE Symposium on Security and Privacy (SP)*, IEEE Computer Society, 2022, 116–133 (cit. on p. 10).

[28] *Owasp top 10: 2021*, 2021. [Online]. Available: https://owasp.org/Top10/ (cit. on p. 10).

[29] W. Koch, A. Chaabane, M. Egele, W. Robertson, and E. Kirda Semi-automated discovery of server-based information oversharing vulnerabilities in android applications, in *Proceedings of the 26th ACM SIGSOFT International Symposium on Software Testing and Analysis*, 2017, 147–157 (cit. on p. 10).

[30] S. Segura, J. A. Parejo, J. Troya, and A. Ruiz-Cortés, Metamorphic testing of restful web apis, *IEEE Transactions on Software Engineering*, vol. 44, no. 11, 1083–1099, 2017 (cit. on p. 10).

[31] P. Godefroid, B.-Y. Huang, and M. Polishchuk Intelligent rest api data fuzzing, in *Proceedings of the 28th ACM Joint Meeting on European Software Engineering Conference and Symposium on the Foundations of Software Engineering*, 2020, 725–736 (cit. on p. 10).

[32] S. Andrica, and G. Candea Warr: A tool for high-fidelity web application record and replay, in *2011 IEEE/IFIP 41st International Conference on Dependable Systems & Networks (DSN)*, IEEE, 2011, 403–410 (cit. on p. 11).

[33] B. Burg, R. Bailey, A. J. Ko, and M. D. Ernst Interactive record/replay for web application debugging, in *Proceedings of the 26th annual ACM symposium on User interface software and technology*, 2013, 473–484 (cit. on p. 11).

[34] S. Flesca, F. Furfaro, and E. Masciari Monitoring web information changes, in *Proceedings International Conference on Information Technology: Coding and Computing*, IEEE, 2001, 421–425 (cit. on p. 11).

[35] Y. Wang, D. J. DeWitt, and J.-Y. Cai X-diff: An effective change detection algorithm for xml documents, in *Proceedings 19th international conference on data engineering (Cat. No. 03CH37405)*, IEEE, 2003, 519–530 (cit. on p. 11).

[36] I. Khoury, R. M. El-Mawas, O. El-Rawas, E. F. Mounayar, and H. Artail, An efficient web page change detection system based on an optimized hungarian algorithm, *IEEE Transactions on Knowledge and Data Engineering*, vol. 19, no. 5, 599–613, 2007 (cit. on p. 11).

[37] M. Hammoudi, G. Rothermel, and A. Stocco Waterfall: An incremental approach for repairing record-replay tests of web applications, in *Proceedings of the 2016 24th ACM SIGSOFT International Symposium on Foundations of Software Engineering*, 2016, 751–762 (cit. on p. 11).

[38] F. Shao, R. Xu, W. Haque, ET AL. Webevo: Taming web application evolution via detecting semantic structure changes, in *Proceedings of the 30th ACM SIGSOFT International Symposium on Software Testing and Analysis*, 2021, 16–28 (cit. on p. 11).

[39] S. Schumilo, C. Aschermann, A. Jemmett, A. Abbasi, and T. Holz, Nyx-net: Network fuzzing with incremental snapshots, *arXiv preprint arXiv:2111.03013*, 2021 (cit. on p. 11).